# AI Literacy as a Key Driver of User Experience in AI-Powered Assessment: Insights from Socratic Mind

Meryem Yilmaz Soylu[a], Jeonghyun Lee[b], Jui-Tse Hung[c], Christopher Zhang Cui[d], David A. Joyner[e],

[a] Center for 21st Century Universities, Georgia Institute of Technology, Atlanta, GA, 30332, USA, meryem@gatech.edu, ORCID: 0000-0003-3080-4686
[b] Center for 21st Century Universities, Georgia Institute of Technology, Atlanta, GA, 30332, USA, jonnalee@gatech.edu, ORCID: 0000-0003-1497-0561
[c] College of Computing, Georgia Institute of Technology, Atlanta, GA, 30332, USA, ruizehung@gatech.edu, ORCID: 0009-0002-4179-2334
[d] Department of Computer Science and Engineering, University of California, San Diego, San Diego, CA 92093, USA, czcui@ucsd.edu, ORCID ID: 0000-0002-0116-634X
[e] College of Computing, Georgia Institute of Technology, Atlanta, GA, 30332, USA, djoyner3@gatech.edu, ORCID ID: 0000-0003-0537-6229

**Corresponding author address**
Center for 21st Century Universities 505 10th St NW, Atlanta, GA 30332, USA

**Corresponding author email address**
meryem@gatech.edu

**Data availability**

Due to the sensitive nature of the questions asked in this study, survey respondents were assured raw data would remain confidential and would not be shared. Anonymized data can be shared upon request.

# AI Literacy as a Key Driver of User Experience in AI-Powered Assessment: Insights from Socratic Mind

## Abstract

As Artificial Intelligence (AI) tools become increasingly embedded in higher education, understanding how students interact with these systems is essential to supporting effective learning experiences. This study investigates how students' AI literacy and prior exposure to AI technologies influence their perceptions of Socratic Mind, an interactive, AI-based formative assessment tool. Drawing on Self-Determination Theory and user experience research, we examine how AI literacy, perceived usability, and satisfaction relate to student engagement and perceived learning effectiveness. Data were collected from 309 undergraduates in Computer Science and Business courses using validated surveys. Partial least squares structural equation modeling revealed that AI literacy—particularly self-efficacy, conceptual understanding, and application skills—significantly predicts usability, satisfaction, and engagement. Usability and satisfaction, in turn, strongly predict perceived learning effectiveness. Prior AI exposure showed no significant effect. These results emphasize that AI literacy—not just exposure—is key to shaping students' experiences with AI tools. To support meaningful engagement, designers should incorporate adaptive guidance and user-centered features that accommodate varying literacy levels. Our findings offer practical guidance for designing an interactive AI-powered assessment systems that foster inclusive, motivating, and effective learning experiences, advancing both theory and practice in interactive learning environments.

# Keywords

1. AI Literacy
2. User Experience (UX)
3. Oral Assessment
4. Human-AI Interaction
5. Socratic Dialogue
6. Self-Determination Theory

# 1. Introduction

Artificial Intelligence (AI) continues to reshape the educational landscape by enabling personalized, adaptive, and interactive learning experiences (Bahroun et al., 2023; Chiu et al., 2024). Recent advances in large language models (LLMs), such as GPT (Brown et al., 2020), and multimodal systems like GPT-4o, have expanded the affordances of educational tools, facilitating dynamic feedback, multimodal interaction, and conversational scaffolding (Achiam et al., 2023; Gris et al., 2023).

One promising application of AI is in assessment, where AI-powered tools such as chatbots and automated feedback systems offer real-time interaction, adaptive guidance, and opportunities for learner reflection and autonomy (Wu & Yu, 2024; Yin et al., 2021). Meta-analytic evidence shows that such tools can enhance engagement, motivation, and academic performance (Chiu et al., 2024; Debets et al., 2025). However, their impact varies across learners, influenced by students' AI literacy and how they experience the interactive qualities of these systems (Walter, 2024; Xu, 2024).

This study investigates undergraduate students' experiences with Socratic Mind, an AI-powered interactive oral assessment tool. Specifically, we examine how AI literacy and user experience (UX) variables (usability, satisfaction) predict engagement and perceived learning effectiveness, and whether prior AI exposure or demographics moderate these outcomes. By doing so, we contribute to the literature on interactive learning environments by clarifying how AI literacy shapes learners' engagement with AI-mediated assessments. The findings offer actionable insights for designing AI-powered tools that foster effective and motivating learning experiences.

## 1.1. Research Questions

1. How do students' AI literacy, perceived usability, and satisfaction influence their engagement and perceived learning effectiveness with an AI-powered oral assessment tool?
2. How do students with higher and lower levels of AI literacy differ in their perceptions of the tool's effectiveness, engagement, satisfaction, and usability?
3. How does prior exposure to AI technologies influence students' perceptions of the tool?
4. How do students' demographic and academic characteristics relate to their AI literacy, AI exposure, and perceptions of the tool?

## 2. Literature Review

### 2.1. AI-Powered Oral Assessment

Traditional assessments such as multiple-choice exams and written assignments emphasize rote recall and often fail to probe students' conceptual understanding (Delson et al., 2022; Sabin et al., 2021). The rise of LLMs further complicates these formats, raising concerns about academic integrity and reducing the diagnostic value of static testing (Gardner & Giordano, 2023; Yanai & Lercher, 2024).

Oral assessments offer a promising alternative by enabling real-time dialogue, probing reasoning, and providing immediate feedback (Qi et al., 2023; Sabin et al., 2021). They promote metacognitive awareness, engagement, and deeper learning (Reckinger & Reckinger, 2022), but their scalability has been limited. AI-powered tools like Socratic Mind (Hung et al., 2024) address this challenge by simulating scalable, interactive Socratic dialogue.

## 2.2. AI Literacy

AI literacy is a multidimensional competency that enables individuals to understand, use, and critically evaluate AI systems (Long & Magerko, 2020; Lintner, 2024). Core dimensions include conceptual understanding, practical application, and critical evaluation (Carolus et al., 2023; Wang et al., 2023).

Higher AI literacy equips students to calibrate trust, evaluate feedback, and engage more productively with AI tools (Huang & Ball, 2024; Klingbeil et al., 2024). AI-literate students typically report greater satisfaction, usability, and perceived learning support in AI-based environments (Al-Abdullatif & Alsubaie, 2024; Xiao et al., 2024). However, prior exposure to AI alone is not a reliable predictor of effective engagement; the depth and quality of AI literacy matter more than frequency of use (Wu & Yu, 2024).

Despite growing interest in AI literacy, little empirical work connects it to students' experiences with interactive AI tools—particularly in assessment contexts. This study addresses that gap by investigating how these factors influence learners' perceived engagement, usability, satisfaction, and perceived learning effectiveness when interacting with an AI-powered oral assessment system.

## 2.3. Self-Determination Theory

Self-Determination Theory (SDT) (Deci et al., 1985; Ryan & Deci, 2000) provides a well-established lens for understanding how technology-mediated environments shape learner motivation. When learners' needs for autonomy, competence, and relatedness are supported, they exhibit more volitional engagement, persistence, and deeper learning outcomes (Chiu et al., 2021; Ryan & Deci, 2017; Yang et al., 2025). These needs are mutually reinforcing: for example, feeling competent enables greater autonomy and fosters relatedness through social affirmation (Wu, 2024; Ruzek et al., 2016; Salikhova et al., 2020).

In interactive learning environments, where learners engage not only with instructional content but also with digital tools, the degree to which these tools support basic psychological needs strongly influences UX and learning outcomes (Hsu et al., 2019; Jeno et al., 2019; Nikou & Economides, 2018). Applying SDT in this context offers a useful framework for understanding students' experiences with AI-powered assessment.

In this study, we conceptually link key UX variables to the three psychological needs. Perceived usability is aligned with autonomy: intuitive, responsive systems help students feel in control of their interaction. AI literacy corresponds to competence, reflecting students' ability to understand, navigate, and benefit from AI systems. Satisfaction relates to relatedness: AI-generated feedback that feels constructive and personalized can foster a sense of social presence, even in AI-mediated settings.

This SDT-based framework helps clarify how interactive features of AI tools, together with students' competencies, shape engagement and perceived learning effectiveness in AI-powered assessments.

Taken together, these constructs inform a comprehensive investigation of how students engage with AI-powered oral assessment tools. By examining how AI literacy and UX variables shape engagement and perceived learning effectiveness, this study contributes new insights into the design of interactive AI technologies that support learner motivation and success. In doing so, it advances both theory and practice in the development of inclusive, learner-centered interactive learning environments.

## 3. Methodology

### 3.1. Participants

Only students who completed at least one full interaction with Socratic Mind and responded to the full UX survey were included in the analysis. A total of 338 undergraduate students completed both the activity and the survey; of these, 309 responses were complete and valid for analysis. Participants were drawn from foundational computer science courses (Introduction to Computing and Systems and Networks; n = 230) and a junior-level business course (Foundations of Strategy; n = 86).

The sample was demographically diverse. Among those reporting gender, 149 identified as male (47.2%) and 154 as female (48.7%). Ethnicity was reported as African, Asian, Pacific American, or Islander (36.4%), Caucasian (25.9%), Latina/Latino (5.1%), and Other (16.5%), with some participants not disclosing. Academic standing included high school dual enrollment students (33.5%), freshmen (9.8%), sophomores (15.5%), juniors (22.2%), seniors (15.5%), and one master's student (0.3%).

Participants represented a range of academic units, including Business (36.7%), Computing (24.1%), Science (12.3%), Engineering (8.9%), Design (0.6%), and Liberal Arts (0.3%), with 14.6% undisclosed. Regarding language background, 81.0% reported English as their native language, 14.2% reported another native language, and 2.2% preferred not to answer.

### 3.2. Design & Instruments

#### 3.2.1. AI-Powered Interactive Oral Assessment Tool: Socratic Mind

Socratic Mind is an interactive AI-based oral assessment tool that supports learner-centered, dialogic engagement. It employs LLMs and automatic speech recognition (ASR) to simulate dynamic, spoken Socratic dialogue. The system prompts students to verbalize reasoning, reflect on understanding, and iteratively explore subject matter concepts through adaptive questioning.

The tool applies principles of Socratic questioning—eliciting clarification, probing assumptions, and encouraging evidence-based reasoning—to foster deeper cognitive engagement. Its design aligns with active learning principles and Webb's Depth of Knowledge framework (Webb, 2002). Unlike traditional written assessments, Socratic Mind emphasizes spontaneous, spoken articulation to externalize thought processes and uncover misconceptions and prompt learners to clarify their understanding. By engaging in dialogue, students often realize gaps in their knowledge or refine their questions—supporting metacognitive awareness and deeper learning.

Students interact with the system primarily via voice input. Pre-configured questions initiate each dialogue, and adaptive follow-up questions are generated in text format based on student responses. Students can progress through the dialogue or end it at any point. Upon completion, Socratic Mind provides automated feedback, including a conversation summary highlighting strengths and areas for improvement.

Instructional features include real-time voice recognition, feedback summaries, dialogue histories, and progress tracking. Instructors configure practice quizzes, define question prompts and desired answers, and set adaptive questioning logic to address common misconceptions. An AI-powered question design tool also supports instructors in crafting questions aligned with course content.

Socratic Mind is scalable for digitally mediated learning environments that prioritize interactive, feedback-rich assessment. Its focus on promoting oral reasoning and adaptive feedback supports both formative and summative applications.

To better understand how students experience and engage with this interactive AI-powered environment, we conducted a survey-based study capturing key dimensions of UX, learning perceptions, and AI literacy. The survey included both closed- and open-ended items assessing perceived usability, engagement, satisfaction, learning effectiveness, AI literacy, and relevant demographic and background information.

### 3.2.2. UX Measures

Students' perceptions of Socratic Mind were evaluated using a 15-item UX survey designed to capture three dimensions grounded in SDT (Deci et al., 1985). The usability subscale (5 items) was conceptualized as supporting autonomy, as intuitive and easy-to-navigate systems enhance users' perceived control and volitional engagement. A representative item includes: "I felt confident interacting with Socratic Mind." The perceived learning effectiveness subscale (5 items) assessed students' perceived learning gains, with sample items such as "Socratic Mind helped enhance my understanding of the topic" and "Using Socratic Mind was valuable to my learning experience." The engagement subscale (5 items) measured learners' motivational involvement and perceptions of the responsiveness and relevance of the tool's questions and feedback, exemplified by "Socratic Mind kept me engaged throughout my learning." Responses were rated on a 5-point Likert scale (1 = Strongly Disagree to 5 = Strongly Agree). Internal consistency was strong across all three subscales (α = .87 for usability, α = .87 for perceived learning effectiveness, α = .88 for engagement).

Two additional items assessed user satisfaction, which we interpret as an indirect proxy for the SDT need for relatedness in this context—that is, the extent to which learners felt socially and affectively supported through interaction with the AI system. These items evaluated general satisfaction and willingness to recommend the tool to peers, using a 5-point labeled scale (e.g., very dissatisfied to very satisfied), with acceptable reliability (α = .74).

### 3.2.3. Meta AI Literacy Scale (MAILS)

AI literacy was operationalized using the Meta AI Literacy Scale (MAILS) (Carolus et al., 2023), a validated 34-item self-report instrument designed to measure individuals' perceived knowledge, skills, and confidence in interacting with AI technologies. The scale encompasses seven dimensions: Apply AI (practical usage), Understand AI (conceptual comprehension), Detect AI (awareness of AI-generated content), AI Ethics (ethical and societal considerations), Create AI (ability to develop or contribute to AI tools), AI Self-Efficacy (confidence in engaging with AI), and AI Self-Competency (general perceived competence).

Participants rated each item on an 11-point Likert scale ranging from 0 (not at all pronounced) to 10 (almost perfectly pronounced). Subscales demonstrated excellent internal consistency (α values ranging from .89 to .98). A composite AI literacy score was computed by averaging across subscales, providing a multidimensional indicator of students' self-perceived AI capabilities. Higher scores indicate greater perceived proficiency in engaging with AI technologies, corresponding to competence in the SDT framework, both in terms of conceptual understanding and practical application. The composite AI literacy score provides a holistic indicator of participants' perceived competence in interacting with, evaluating, and creating AI technologies, corresponding to the competence dimension of the SDT framework.

### 3.2.4. Demographics and AI Exposure

Demographic variables were collected to contextualize participants' academic and personal backgrounds. Items included age, gender, academic standing, GPA, college affiliation, ethnicity, native language, and self-reported English proficiency. These variables allowed for subgroup analyses related to UX and AI literacy.

AI exposure was operationalized as participants' self-reported prior experience with generative AI technologies. Items captured both the frequency of AI use and the primary context of that use (e.g., academic vs. everyday life). This information was used to explore whether prior exposure moderated students' perceptions of *Socratic Mind* or influenced levels of AI literacy. Although prior AI exposure does not equate to AI literacy, it provides important contextual information for interpreting variability in UX and literacy outcomes.

### 3.2.5. Procedure

This study received ethical approval from the authors' institutional review board (Protocol #IRB2024-84). Informed consent was obtained from all participants.

Socratic Mind was embedded into course activities across the semester in computer science and business courses as a formative assessment tool. In both contexts, students used Socratic Mind as a practice quiz to check their understanding of key course content and to prepare for actual quizzes and exams. The tool engaged students in AI-guided dialogue and feedback designed to promote reflective learning and reinforce conceptual understanding. Participation was optional and varied by instructor implementation, with all instructors encouraging meaningful interaction with the tool throughout the semester.

At the end of the semester, students were invited to complete a voluntary and anonymous UX survey via email or the learning management system. The survey was administered just before finals week, allowing students to reflect on their full-semester experience with Socratic Mind.

## 4. Results

### 4.1. Exploring How AI Literacy, Usability, and Satisfaction Drive Engagement and Effectiveness

Partial Least Squares Structural Equation Modeling (PLS-SEM) was conducted using SmartPLS 4 (Ringle et al., 2020) to examine how AI literacy (representing competence), usability (autonomy), and satisfaction (relatedness) influence students' perceptions of and experiences with Socratic Mind, particularly in relation to learning engagement and effectiveness (see Figure 1). PLS-SEM was selected due to the complexity of the proposed model, which includes multiple latent constructs and hypothesized pathways, and the study's emphasis on theory building and predictive analysis (Chin et al., 2008). This method is particularly well-suited for exploratory research where the goal is to estimate causal relationships among constructs and assess the predictive relevance of the model. It has been widely applied in studies of technology adoption, decision-making, and user satisfaction in digital environments (Al-Emran et al., 2024; Huang, 2021; Melchor & Julián, 2008; Pavlou & Fygenson, 2006), making it an appropriate analytical approach for examining how AI literacy, usability, and satisfaction influence students' perceptions of learning engagement and effectiveness when using Socratic Mind.

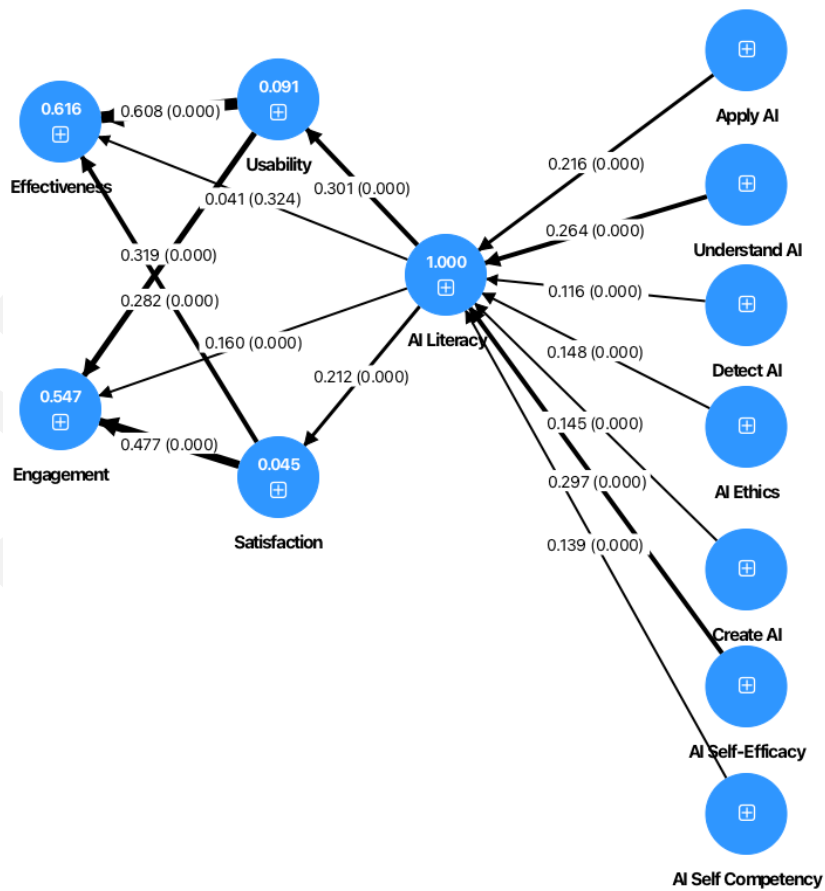

Figure 1. Structural Model

The measurement model demonstrated strong reliability and validity. All constructs showed high internal consistency reliability, with composite reliability values ranging from .739 (Satisfaction) to .982 (Create AI), and Cronbach's alpha values above .70 (see Supplementary Table S1).

Convergent validity was generally supported, as Average Variance Extracted (AVE) values exceeded .50 for all constructs except AI Literacy (AVE = .432). While the AVE for the second-order construct AI Literacy was slightly below the .50 threshold, it was retained due to strong theoretical support and internal consistency (CR = .849). Discriminant validity, assessed via the Heterotrait-Monotrait (HTMT) criterion, was satisfactory as all HTMT values remained below .85 (Henseler et al., 2015) (see Supplementary Table S2).

The formative indicators of AI Literacy—Understand AI ($\beta$ = .264), AI Self-Efficacy ($\beta$ = .297), Apply AI ($\beta$ = .216), and Create AI ($\beta$ = .145)—were all statistically significant ($ps < .001$), supporting the robustness of the construct and emphasizing the importance of these foundational competencies. AI Literacy, conceptualized as a competency construct, had significant positive effects on Usability (aligned with autonomy; $\beta$ = .301, $t$ = 5.093, $p < .001$), Satisfaction (aligned with relatedness; $\beta$ = .212, $t$ = 3.620, $p < .001$), and Engagement ($\beta$ = .160, $t$ = 3.962, $p < .001$), indicating that higher levels of AI Literacy are associated with enhanced UX across multiple dimensions. However, the direct effect of AI Literacy on perceived learning Effectiveness was not statistically significant ($\beta$ = .041, $t$ = 0.987, $p$ = .324), suggesting a possible indirect effect through mediating variables such as Usability and Satisfaction—both of which were strong direct predictors of Effectiveness (see Supplementary Table S3).

The model explained a substantial portion of the variance in Effectiveness ($R^2 = .616$) and Engagement ($R^2 = .547$), with lower variance in Usability ($R^2 = .091$) and Satisfaction ($R^2 = .045$) (see Supplementary Table S4).

Effect sizes ($f^2$) showed that Usability had a large effect on Effectiveness ($f^2 = .743$), and Satisfaction had moderate effects on both Effectiveness and Engagement. Predictors like AI Self-Efficacy, Apply AI, and Understand AI had extremely large effects on AI Literacy (see Supplementary Table S5).

These findings confirm that AI literacy significantly enhances Usability and Satisfaction dimensions and indirectly improves perceived learning Effectiveness and Engagement.

### 4.2. The Role of AI Literacy in Shaping Student Experience

To determine participants' AI literacy level, we used the mean AI literacy score ($M = 8.4$) as a cutoff. Participants with scores below 8.4 were categorized into the low AI literacy group, while those with scores equal to or above 8.4 were categorized into the high AI literacy group. Independent samples t-tests were conducted to compare satisfaction, usability, effectiveness, and engagement ratings between the two groups.

*Table 1*

*Comparison of Satisfaction, Usability, Effectiveness, and Engagement by AI Literacy Level*

| Variable | AI Literacy Level | N | $M$ | SD | $t$(df) | $p$ (2-tailed) | Cohen's $d$ |
|---|---|---|---|---|---|---|---|
| Satisfaction | Low | 143 | 4.22 | 0.74 | -2.87 | .004 | -0.33 |
|  | High | 168 | 4.46 | 0.67 | (288.44) |  |  |
| Usability | Low | 143 | 4.53 | 0.65 | -3.98 | <.001 | -0.47 |
| | High | 168 | 4.78 | 0.38 | (220.80) | | |
| Effectiveness | Low | 143 | 4.49 | 0.62 | -4.05 | <.001 | -0.47 |
| | High | 168 | 4.74 | 0.43 | (245.68) | | |
| Engagement | Low | 143 | 4.11 | 0.76 | -5.83 | <.001 | -0.68 |
| | High | 168 | 4.57 | 0.59 | (264.34) | | |

As shown in Table 1, participants with high AI literacy reported significantly higher ratings across satisfaction, usability, effectiveness, and engagement, compared to those with low AI literacy. Effect sizes ranged from small to medium for satisfaction and usability and approached large for engagement. These results suggest that higher AI literacy, competency, is associated with more favorable perceptions of the AI system across all dimensions, with effect sizes ranging from small to medium for satisfaction and usability and approaching large for engagement.

## 4.3. Examining the Influence of Prior AI Exposure on Student Perceptions of an AI Assessment Tool

To determine participants' level of exposure to an AI tool, we calculated an exposure score for each student by averaging their responses to two items measuring how frequently they use AI at school and in everyday life, both rated on a 5-point Likert scale. The mean exposure score was 2.49, which was consistent with the median. Based on this value, participants were categorized into two groups: those with scores below 2.49 were classified as having low exposure, and those with scores of 2.49 or higher were classified as having high exposure. Independent samples t-tests were then conducted to compare satisfaction, usability, effectiveness, and engagement ratings between the two groups.

*Table 2*

*Comparison of Satisfaction, Usability, Effectiveness, and Engagement by AI Exposure Level*

| Variable | Exposure Level | N | M | SD | t(df) | p (2-tailed) | Cohen's d |
|---|---|---|---|---|---|---|---|
| Satisfaction | Low | 172 | 4.32 | 0.74 | -0.91 | .362 | -0.11 |
| | High | 127 | 4.39 | 0.65 | (287.73) | | |
| Usability | Low | 172 | 4.66 | 0.54 | -1.09 | .278 | -0.12 |
| | High | 127 | 4.72 | 0.43 | (295.57) | | |
| Effectiveness | Low | 172 | 4.63 | 0.55 | -0.47 | .636 | -0.05 |
| | High | 127 | 4.66 | 0.44 | (295.53) | | |
| Engagement | Low | 172 | 4.33 | 0.73 | -1.12 | .262 | -0.13 |
| | High | 127 | 4.42 | 0.66 | (284.50) | | |

Note. Low exposure group = Exposure score < 2.49. High exposure group = Exposure score ≥ 2.49.

Results showed no statistically significant differences between the low- and high-exposure groups in terms of satisfaction, usability, effectiveness, or engagement (Table 2), indicating that the level of AI exposure did not significantly influence participants' perceptions of their experience with Socratic Mind.

## 4.4. How Demographic and Academic Factors Relate to AI Literacy and Socratic Mind Experience

Descriptive and inferential statistical analyses were conducted to examine whether students who reported prior use of generative AI tools differed significantly from those who did not across various outcome measures. Generative AI (GAI) tool usage was measured with a single binary item asking participants whether they had used any generative AI tools (e.g., ChatGPT, DALL-E, Grammarly) before engaging with Socratic Mind. Table 3 presents the comparison of perception and experience ratings and AI literacy subscale scores based on students' reported GAI tool usage.

*Table 3*

*Comparison of Perceptions and AI Literacy Subscale Scores by Generative AI Tool Usage*

| Variable | AI Tool Use | N | M | SD | t(df) | p (2-tailed) | Cohen's d |
|---|---|---|---|---|---|---|---|
| Usability | No | 89 | 4.56 | 0.64 | -1.99 | .048 | -0.28 |
|  | Yes | 216 | 4.71 | 0.48 | (130.63) |  |  |
| Effectiveness | No | 89 | 4.57 | 0.61 | -0.90 | .368 | -0.12 |
|  | Yes | 216 | 4.64 | 0.52 | (142.90) |  |  |
| Engagement | No | 89 | 4.31 | 0.78 | -0.52 | .604 | -0.07 |
|  | Yes | 216 | 4.36 | 0.69 | (146.91) |  |  |
| Satisfaction | No | 89 | 4.30 | 0.73 | -0.75 | .457 | -0.10 |
|  | Yes | 216 | 4.37 | 0.71 | (159.56) |  |  |
| AI Literacy | No | 89 | 8.27 | 1.76 | -0.73 | .468 | -0.10 |
|  | Yes | 216 | 8.42 | 1.34 | (131.95) |  |  |
| AI Exposure | No | 84 | 2.14 | 1.08 | -3.65 | <.001 | -0.50 |
|  | Yes | 214 | 2.63 | 0.94 | (135.41) |  |  |
| Apply AI | No | 89 | 9.05 | 1.93 | -2.72 | .007 | -0.40 |
|  | Yes | 216 | 9.66 | 1.34 | (124.47) |  |  |
| Understand AI | No | 89 | 8.34 | 2.10 | -1.35 | .179 | -0.19 |
|  | Yes | 216 | 8.68 | 1.69 | (137.00) |  |  |
| Detect AI | No | 89 | 8.39 | 2.13 | -2.11 | .037 | -0.30 |
|  | Yes | 216 | 8.91 | 1.53 | (127.39) |  |  |
| Create AI | No | 89 | 6.55 | 3.17 | 2.41 | .017 | 0.32 |
| | Yes | 216 | 5.61 | 2.88 | (150.93) | | |
| AI Self-Efficacy | No | 89 | 7.93 | 2.42 | -1.59 | .114 | -0.22 |
| | Yes | 216 | 8.39 | 1.95 | (137.62) | | |
| AI Self-Competency | No | 89 | 9.38 | 1.63 | 0.44 | .659 | 0.06 |
| | Yes | 216 | 9.29 | 1.50 | (153.32) | | |

Users of GAI tools reported significantly higher average usability scores compared to non-users. Similarly, GAI users reported significantly greater AI exposure than non-users. Significant differences were also found for the ability to apply AI, with GAI users scoring higher than non-users. GAI users also scored higher in their ability to detect AI-generated content. In contrast, non-users reported significantly higher scores in their ability to create AI compared to users. No significant differences were found between groups for perceived effectiveness, engagement, satisfaction, AI literacy, understanding of AI, AI self-efficacy, or AI self-competency (all $p$s > .05). These results suggest that while GAI tool users tend to report greater usability and confidence in applying and detecting AI, non-users may feel more capable when it comes to creating AI.

An Analysis Of VAriance (ANOVA) was conducted to examine whether students' academic standing (e.g., high school, freshman, sophomore, etc.) was associated with differences across various outcomes. No statistically significant differences were found for average usability, $F(6, 301) = 1.16$, $p = .325$; effectiveness, $F(6, 301) = 0.69$, $p = .659$; engagement, $F(6, 301) = 1.30$, $p = .259$; ability to apply AI skills, $F(6, 301) = 1.32$, $p = .246$; understanding of AI, $F(6, 301) = 1.71$, $p = .119$; ability to detect AI-generated content, $F(6, 300) = 1.12$, $p = .352$; ability to create AI content, $F(6, 300) = 0.37$, $p = .897$; AI self-efficacy, $F(6, 300) = 1.35$, $p = .235$; or overall satisfaction, $F(6, 301) = 0.49$, $p = .813$. However, significant differences emerged for AI self-competency, $F(6, 301) = 3.06$, $p = .006$, and AI exposure, $F(6, 292) = 3.25$, $p = .004$.

Post hoc Bonferroni comparisons revealed that high school students reported significantly lower AI exposure scores than both juniors ($p = .017$) and seniors ($p = .009$). This dimension reflects students' perceived confidence in exploring and experimenting with AI tools—a skill likely strengthened through increased access to AI-integrated coursework and hands-on experiences. Seniors also reported significantly higher AI self-competency than high school students ($p = .006$). This broader dimension encompasses not just technical proficiency but also the ability to apply AI tools meaningfully in real-world contexts. These findings suggest that more advanced students may have had greater exposure to AI applications through academic projects, internships, or extracurricular activities. While other AI literacy dimensions did not reach statistical significance, emerging trends suggest a developmental trajectory in which AI literacy strengthens with academic progression.

An independent samples *t*-test was conducted to examine gender differences in perceived usability, effectiveness, engagement, AI-related competencies, and satisfaction. (see Supplementary Table S5). No significant gender differences emerged in usability, effectiveness, engagement, AI exposure, satisfaction, or ability to apply AI. However, males scored significantly higher on overall AI literacy, understanding of AI, and the ability to create AI content. Additionally, males reported greater AI self-efficacy. While differences in detecting AI-generated content and AI self-competency did not reach conventional levels of significance, they approached marginal significance ($p$s = .081 and .094, respectively), with males scoring higher. These results suggest that, although general learning experience variables are consistent across genders, male participants reported significantly higher confidence in several AI-related dimensions, especially understanding, creating, and self-efficacy.

Independent samples *t*-tests were also conducted to examine whether participants' native language status (native English speakers vs. non-native English speakers) influenced perceptions of usability, engagement, satisfaction, and AI-related competencies. No significant differences were observed for usability, $t(63.37) = -0.48$, $p = .631$; effectiveness, $t(59.58) = -0.24$, $p = .810$; engagement, $t(59.89) = -0.33$, $p = .743$; satisfaction, $t(53.44) = -0.12$, $p = .901$; ability to apply AI, $t(62.74) = -0.77$, $p = .444$; or AI self-competency, $t(60.06) = -0.13$, $p = .900$. However, non-native English speakers reported significantly greater AI exposure ($M = 2.79$, $SD = 1.22$) than native speakers ($M = 2.43$, $SD = 0.96$), $t(291) = -2.15$, $p = .032$, and a marginally higher understanding of AI, $t(61.84) = -1.87$, $p = .066$. They also reported slightly higher overall AI literacy scores ($M = 8.71$, $SD = 1.38$) compared to native speakers ($M = 8.33$, $SD = 1.47$), although the difference approached but did not reach statistical significance, $t(62.92) = -1.69$, $p = .096$. No significant differences were observed for the ability to detect or create AI content or for AI self-efficacy. These results suggest that while general learning experience outcomes are similar across language groups, non-native English speakers may feel more confident or have had more exposure to AI tools and concepts.

## 5. Discussion

This study examined how undergraduate students' AI literacy and related competencies shaped their UXs with an AI-powered oral assessment tool, *Socratic Mind*. Using PLS-SEM and a series of group comparisons, we found that AI literacy played a central role in predicting usability, satisfaction, engagement, and perceived learning effectiveness. Specifically, foundational competencies such as AI self-efficacy, application skills, and understanding of AI significantly contributed to students' overall AI literacy, which in turn enhanced their UX. Although AI literacy did not directly predict perceived effectiveness, its influence was mediated through usability and satisfaction.

As AI technologies introduce new layers of complexity to digital interaction, distinct from traditional tools, understanding the multifaceted effects of AI literacy becomes increasingly important. Such knowledge can guide the design of AI tools and learning strategies that promote positive learner outcomes while mitigating potential challenges (Markus et al., 2024; Pinski & Benlian, 2024). Prior research supports this perspective: AI literacy—including conceptual knowledge, applied skills, and self-efficacy—has been linked to improved perceptions of system usability, satisfaction, and engagement (Lee & Park, 2023; Q. Zhang et al., 2025). For example, Lee and Park (2023) found that college students with higher ChatGPT literacy reported significantly greater satisfaction with the tool, regardless of their reasons for using it. Similarly, Q. Zhang et al. (2025) demonstrated that English as foreign language students with stronger AI literacy experienced higher self-efficacy, lower anxiety, and increased willingness to participate in AI-assisted speaking activities.

However, our findings underscore that competence alone is not sufficient to explain students' perceptions of learning effectiveness and engagement. Evidence is emerging that user-centered design and satisfaction are critical mediators in translating AI literacy into positive learning outcomes. In other words, possessing AI knowledge does not automatically lead to enhanced learning unless the tool is perceived as usable and satisfying (Alshammari & Babu, 2025; Jeilani

& Abubakar, 2025). This highlights the importance of designing intuitive, responsive systems that promote not only cognitive engagement but also user comfort and motivation.

Collectively, these findings converge on a key insight: usability and satisfaction serve as crucial intermediary variables. While high AI literacy may create the foundation for productive engagement with AI tools, it is the ease of use and the user's resulting satisfaction that ultimately translate literacy into perceived learning effectiveness.

Although students with higher levels of AI literacy consistently reported more positive UXs, our findings revealed that prior AI exposure alone did not significantly influence perceptions of the tool. This suggests that the frequency or familiarity of past AI use is less important than the depth of knowledge and confidence developed through meaningful engagement with AI technologies. For instance, W. Zhang et al. (2025) reported that students who had previously interacted with AI were no more or less satisfied than those encountering it for the first time. This finding highlights that exposure, without accompanying literacy or positive disposition, may not significantly affect engagement or satisfaction. Further supporting this perspective, Bewersdorff et al. (2025) identified distinct user profiles among university students. One such group termed "Pragmatic Observers" demonstrated moderate interest and occasional AI use but exhibited low AI literacy and confidence. Despite prior exposure, members of this group reported low engagement and self-efficacy when interacting with AI tools. These findings reinforce a critical insight: prior AI experience alone does not predict user experience outcomes. Instead, the quality and depth of that experience, particularly whether it fosters competence and trust, make a substantial contribution to shaping UX. Simply having used an AI tool does not necessarily lead to greater satisfaction, usability, or learning gains unless it is accompanied by deeper understanding and a positive, informed mindset toward the technology.

Our findings also indicated that demographic differences were generally modest, though some variability emerged by gender and academic standing in specific AI literacy subcomponents. This aligns with recent studies such as Asio (2024), who surveyed over 1,000 college students and reported moderate differences in AI literacy, self-efficacy, and self-competence based on college major, year level, and gender. Male students and those in advanced academic years tended to score slightly higher in literacy and confidence, although the overall levels were categorized as "somewhat literate" across groups. Similarly, AI self-efficacy showed minor variation by age and gender, with older or male students reporting slightly greater confidence. These trends were characterized as moderate rather than substantial. Additional studies support this nuanced view. Bewersdorff et al. (2025), for instance, identified student profiles such as "AI Advocates," who had high AI literacy and confidence and were predominantly male and enrolled in engineering or technical disciplines. In contrast, their "Cautious Critics" group—characterized by lower confidence and greater skepticism—had a female majority and more students from humanities backgrounds. Yet, not all aspects of UX vary by demographics. W. Zhang et al. (2025), for example, found no significant gender differences in satisfaction with AI learning tools among high school students. Collectively, these findings suggest that while academic level and gender may influence AI literacy and confidence to some extent, they are not strong determinants of UX outcomes. Promoting AI readiness across diverse learner groups is essential to ensuring equitable and effective use of AI technologies in education.

Our study also contributes to the application of SDT (Deci et al., 1985) in AI-mediated learning environments. By aligning usability, satisfaction, and AI literacy with the psychological needs of autonomy, relatedness, and competence, respectively, we illustrate how AI tools can support or hinder student motivation. Students with higher AI literacy reported greater usability and satisfaction—suggesting that feeling competent in using AI contributes to a sense of control (autonomy) and feeling supported (relatedness). These findings are consistent with prior research showing that digital learning tools that meet users' basic psychological needs foster higher engagement and learning effectiveness (Jeno et al., 2019; Nikou & Economides, 2018). For instance, a usable AI tool enhances autonomy by allowing students to navigate it independently without frustration. Literacy fosters competence, making students feel capable and skilled. Satisfaction, often linked to responsive or personalized feedback, simulates relatedness—even in the absence of human interaction—by creating a sense of social presence.

HCI experts have emphasized that technologies designed to support autonomy, competence, and relatedness tend to generate higher user satisfaction and engagement (Kohler, 2022; Peters et al., 2018). Our findings reinforce this SDT-aligned perspective: usability, literacy, and satisfaction each fulfill a fundamental human motivator, which in turn drives engagement and perceived learning effectiveness. This framework helps explain the mediation effects observed in our model—namely, how the impact of AI literacy on effectiveness is channeled through usability and satisfaction. By meeting learners' basic psychological needs, well-designed AI tools can transform competence into meaningful educational outcomes.

This study contributes to the growing body of knowledge on the design of interactive learning environments that leverage AI technologies. While prior work has emphasized the potential of AI to support assessment and feedback, our findings highlight the critical role of AI literacy in shaping how students engage with and benefit from such environments. These insights extend beyond the specific context of oral assessment to inform the broader design of interactive AI-powered learning environments where adaptive dialogue, user feedback, and learner autonomy are central features.

By demonstrating how usability and satisfaction mediate the relationship between AI literacy and learning effectiveness, this work offers actionable guidance for designing interactive systems that are not only technically robust but also pedagogically effective and inclusive. As conversational AI and adaptive feedback mechanisms become increasingly integrated into learning environments, aligning system design with learners' competencies and psychological needs becomes vital for maximizing educational impact.

This study also adds empirical support for the MAILS (Carolus et al., 2023) in the context of oral assessment, highlighting AI self-efficacy, Apply AI, and Understand AI as especially strong contributors to students' perceived usability, satisfaction, and engagement. These findings align with prior frameworks that emphasize applied, reflective, and ethical dimensions of AI literacy (Long & Magerko, 2020; Wang et al., 2023), and provide a pathway for educators and tool developers to translate theoretical constructs into actionable design and instructional strategies.

Our findings offer several practical implications for improving learner experience with AI-powered tools like *Socratic Mind*. First, developers should consider embedding adaptive supports that respond to users' AI literacy profiles. For instance, scaffolding complex interactions, offering optional guided walkthroughs, or simplifying technical language may improve usability and satisfaction for users with lower literacy levels—enhancing both inclusion and impact (Zawacki-Richter et al., 2019). Second, interface and feedback design should explicitly target the core psychological needs identified by SDT. Enhancing usability through intuitive design fosters autonomy; increasing opportunities for students to successfully apply their AI skills builds competence; and personalized, encouraging feedback fosters relatedness. These elements work together to promote deeper engagement and more effective learning.

Third, our results suggest that UX is shaped not only by competence but also by how that competence is activated through supportive, responsive tool design. Developers should prioritize features like real-time, explainable feedback and transparency mechanisms that increase trust and reduce uncertainty—key elements of effective interactive learning design in AI-mediated environments—particularly important for students who may be unfamiliar or hesitant with AI technologies (Roll & Wylie, 2016). Fourth, instructors should not assume that prior exposure to AI translates to readiness. Educational programs should therefore integrate explicit instruction in core AI competencies—especially those shown to predict user satisfaction and engagement—alongside opportunities for students to reflect on and evaluate their interactions with AI systems.

Finally, equity must remain central. While gender and academic standing showed only modest differences, these variations suggest the need for proactive strategies to support students from less technically aligned disciplines or with lower confidence in AI. Building AI literacy across diverse student populations through formal instruction, co-curricular training, and embedded tool supports can help ensure more equitable and empowering AI learning environments. When AI tools are thoughtfully aligned with learners' competencies and psychological needs, they can enhance student motivation, autonomy, and success.

# 6. Limitations and Future Research

This study has several limitations that warrant consideration. First, the reliance on self-report measures introduces potential biases such as social desirability and inaccurate recall. Although the constructs demonstrated strong internal consistency and theoretical alignment, future research should incorporate behavioral data (e.g., interaction logs, usage patterns, performance analytics) to triangulate and validate students' self-perceptions (Arizmendi et al., 2023; Salehian Kia et al., 2021). Second, the study was conducted at a single research-intensive institution with a relatively high proportion of students in computing and business disciplines. As such, the findings may not generalize to institutions with different student demographics, cultural contexts, or curricular orientations. Broader multi-institutional or cross-cultural studies could help establish the robustness of the observed relationships across diverse educational settings.

Third, the cross-sectional design precludes strong causal inferences. While PLS-SEM supports the modeling of hypothesized relationships, longitudinal research is needed to investigate how AI literacy develops over time and how sustained use of AI tools influences engagement and learning outcomes. Fourth, while the study explored demographic differences, the nuanced role of intersectional factors (e.g., gender, major, age, and digital fluency) in shaping AI experiences remains underexplored.

Future research should also assess the impact of targeted AI literacy interventions—particularly those emphasizing self-efficacy, application, and understanding—on learners' engagement, satisfaction, and performance with AI tools. Finally, qualitative studies, such as interviews or think-aloud protocols, could yield richer insights into students' trust, perceived agency, and sense of connection when interacting with AI-powered assessments, especially among underrepresented or digitally less confident student populations.

## 7. Conclusion

As artificial intelligence becomes more deeply embedded in educational systems, identifying the competencies that enable students to engage effectively with these technologies is increasingly important. This study offers empirical support for the central role of AI literacy, especially the dimensions of self-efficacy, application skills, and conceptual understanding, in shaping learners' experiences with interactive AI-based assessment tools. In contrast, prior AI exposure alone was not a significant predictor of usability, satisfaction, engagement, or perceived effectiveness, highlighting the distinction between familiarity and functional competence.

Our findings demonstrate that AI literacy exerts its strongest effects indirectly, through its influence on usability and satisfaction, two key mediators of perceived learning effectiveness. These results are consistent with Self-Determination Theory, suggesting that tools designed to support users' autonomy, competence, and relatedness foster more engaging and effective learning environments. They underscore the need for AI tools that provide responsive, transparent feedback tailored to users' literacy levels.

For researchers and practitioners in human-computer interaction, educational technology, and the design of interactive learning environments, these findings point to two critical priorities: advancing the development of AI literacy across diverse student populations and ensuring that AI systems are thoughtfully designed to support UX and psychological needs. Together, these efforts can help ensure that AI-enhanced learning environments are not only technologically sophisticated, but also pedagogically inclusive and human-centered.

# 8. References

Achiam, J., Adler, S., Agarwal, S., Ahmad, L., Akkaya, I., Aleman, F. L., Almeida, D., Altenschmidt, J., Altman, S., & Anadkat, S. (2023). Gpt-4 technical report. *arXiv preprint arXiv:2303.08774*.

Al-Abdullatif, A. M., & Alsubaie, M. A. (2024). ChatGPT in learning: Assessing students' use intentions through the lens of perceived value and the influence of AI literacy. *Behavioral Sciences*, *14*(9), 845.

Al-Emran, M., Al-Sharafi, M. A., Foroughi, B., Iranmanesh, M., Alsharida, R. A., Al-Qaysi, N., & Ali, N. a. (2024). Evaluating the barriers affecting cybersecurity behavior in the Metaverse using PLS-SEM and fuzzy sets (fsQCA). *Computers in Human Behavior*, *159*, 108315.

Alshammari, S. H., & Babu, E. (2025). The mediating role of satisfaction in the relationship between perceived usefulness, perceived ease of use and students' behavioural intention to use ChatGPT. *Scientific Reports*, *15*(1), 7169.

Arizmendi, C. J., Bernacki, M. L., Raković, M., Plumley, R. D., Urban, C. J., Panter, A., Greene, J. A., & Gates, K. M. (2023). Predicting student outcomes using digital logs of learning behaviors: Review, current standards, and suggestions for future work. *Behavior research methods*, *55*(6), 3026-3054.

Asio, J. M. R. (2024). AI literacy, self-efficacy, and self-competence among college students: variances and interrelationships among variables. *MOJES: Malaysian Online Journal of Educational Sciences*, *12*(3), 44-60.

Bahroun, Z., Anane, C., Ahmed, V., & Zacca, A. (2023). Transforming education: A comprehensive review of generative artificial intelligence in educational settings through bibliometric and content analysis. *Sustainability*, *15*(17), 12983.

Bewersdorff, A., Hornberger, M., Nerdel, C., & Schiff, D. S. (2025). AI advocates and cautious critics: How AI attitudes, AI interest, use of AI, and AI literacy build university students' AI self-efficacy. *Computers and Education: Artificial Intelligence*, *8*, 100340.

Brown, T., Mann, B., Ryder, N., Subbiah, M., Kaplan, J. D., Dhariwal, P., Neelakantan, A., Shyam, P., Sastry, G., & Askell, A. (2020). Language models are few-shot learners. *Advances in Neural Information Processing Systems*, *33*, 1877-1901.

Carolus, A., Koch, M. J., Straka, S., Latoschik, M. E., & Wienrich, C. (2023). MAILS-Meta AI literacy scale: Development and testing of an AI literacy questionnaire based on well-founded competency models and psychological change-and meta-competencies. *Computers in Human Behavior: Artificial Humans*, *1*(2), 100014.

Chin, W. W., Peterson, R. A., & Brown, S. P. (2008). Structural equation modeling in marketing: Some practical reminders. *Journal of marketing theory and practice*, *16*(4), 287-298.

Chiu, T. K., Moorhouse, B. L., Chai, C. S., & Ismailov, M. (2024). Teacher support and student motivation to learn with Artificial Intelligence (AI) based chatbot. *Interactive Learning Environments*, *32*(7), 3240-3256.

Debets, T., Banihashem, S. K., Joosten-Ten Brinke, D., Vos, T. E., de Buy Wenniger, G. M., & Camp, G. (2025). Chatbots in Education: A Systematic Review of Objectives, Underlying Technology and Theory, Evaluation Criteria, and Impacts. *Computers & Education*, 105323.

Deci, E. L., Ryan, R. M., Deci, E. L., & Ryan, R. M. (1985). Conceptualizations of intrinsic motivation and self-determination. *Intrinsic motivation and self-determination in human behavior*, 11-40.

Delson, N., Baghdadchi, S., Ghazinejad, M., Lubarda, M., Minnes, M., Phan, A., Schurgers, C., & Qi, H. (2022). Can oral exams increase student performance and motivation? Proceedings ASEE annual conference,

DigitalEducationCouncil. (2024). *Global AI Student Survey 2024. Digital Education Council*. https://www.digitaleducationcouncil.com/post/digital-education-council-global-ai-student-survey-2024

Gardner, D. E., & Giordano, A. N. (2023). The challenges and value of undergraduate oral exams in the physical chemistry classroom: A useful tool in the assessment toolbox. *Journal of Chemical Education*, *100*(5), 1705-1709.

Gris, L. R. S., Marcacini, R., Junior, A. C., Casanova, E., Soares, A., & Aluísio, S. M. (2023). Evaluating OpenAI's Whisper ASR for Punctuation Prediction and Topic Modeling of life histories of the Museum of the Person. *arXiv preprint arXiv:2305.14580*.

Henseler, J., Ringle, C. M., & Sarstedt, M. (2015). A new criterion for assessing discriminant validity in variance-based structural equation modeling. *Journal of the academy of marketing science*, *43*, 115-135.

Huang, C.-H. (2021). Using PLS-SEM model to explore the influencing factors of learning satisfaction in blended learning. *Education Sciences*, *11*(5), 249.

Huang, K. T., & Ball, C. (2024). The Influence of AI Literacy on User's Trust in AI in Practical Scenarios: A Digital Divide Pilot Study. *Proceedings of the Association for Information Science and Technology*, *61*(1), 937-939.

Hung, J.-T., Cui, C., Popescu, D. M., Chatterjee, S., & Starner, T. (2024). Socratic Mind: Scalable Oral Assessment Powered By AI. Proceedings of the Eleventh ACM Conference on Learning@ Scale,

Jeilani, A., & Abubakar, S. (2025). Perceived institutional support and its effects on student perceptions of AI learning in higher education: the role of mediating perceived learning outcomes and moderating technology self-efficacy [Perspective]. *Frontiers in Education*, *Volume 10 - 2025*. https://doi.org/10.3389/feduc.2025.1548900

Jeno, L. M., Adachi, P. J., Grytnes, J. A., Vandvik, V., & Deci, E. L. (2019). The effects of m-learning on motivation, achievement and well-being: A Self-Determination Theory approach. *British Journal of Educational Technology*, *50*(2), 669-683.

Jin, Y., Yang, K., Yan, L., Echeverria, V., Zhao, L., Alfredo, R., Milesi, M., Fan, J. X., Li, X., & Gasevic, D. (2025). Chatting with a learning analytics dashboard: The role of generative AI literacy on learner interaction with conventional and scaffolding chatbots. Proceedings of the 15th International Learning Analytics and Knowledge Conference,

Klingbeil, A., Grützner, C., & Schreck, P. (2024). Trust and reliance on AI—An experimental study on the extent and costs of overreliance on AI. *Computers in Human Behavior*, *160*, 108352.

Kohler, T. (2022). *Autonomy, Relatedness, and Competence in UX Design*. https://www.nngroup.com/articles/autonomy-relatedness-competence/

Laupichler, M. C., Aster, A., Schirch, J., & Raupach, T. (2022). Artificial intelligence literacy in higher and adult education: A scoping literature review. *Computers and Education: Artificial Intelligence*, *3*, 100101.

Lee, S., & Park, G. (2023). Exploring the impact of ChatGPT literacy on user satisfaction: The mediating role of user motivations. *Cyberpsychology, Behavior, and Social Networking*, *26*(12), 913-918.

Li, Y., Wu, B., Huang, Y., & Luan, S. (2024). Developing trustworthy artificial intelligence: insights from research on interpersonal, human-automation, and human-AI trust. *Frontiers in psychology*, *15*, 1382693.

Lintner, T. (2024). A systematic review of AI literacy scales. *npj Science of Learning*, *9*(1), 50.

Long, D., & Magerko, B. (2020). What is AI literacy? Competencies and design considerations. Proceedings of the 2020 CHI conference on human factors in computing systems,

Markus, A., Pfister, J., Carolus, A., Hotho, A., & Wienrich, C. (2024). Effects of AI understanding-training on AI literacy, usage, self-determined interactions, and anthropomorphization with voice assistants. *Computers and Education Open*, *6*, 100176.

Melchor, M. Q., & Julián, C. P. (2008). The impact of the human element in the information systems quality for decision making and user satisfaction. *Journal of Computer Information Systems*, *48*(2), 44-52.

Nikou, S. A., & Economides, A. A. (2018). Mobile-Based micro-Learning and Assessment: Impact on learning performance and motivation of high school students. *Journal of Computer Assisted Learning*, *34*(3), 269-278.

Pavlou, P. A., & Fygenson, M. (2006). Understanding and predicting electronic commerce adoption: An extension of the theory of planned behavior. *MIS quarterly*, 115-143.

Peters, D., Calvo, R. A., & Ryan, R. M. (2018). Designing for motivation, engagement and wellbeing in digital experience. *Frontiers in psychology*, *9*, 300159.

Pinski, M., & Benlian, A. (2024). AI literacy for users–A comprehensive review and future research directions of learning methods, components, and effects. *Computers in Human Behavior: Artificial Humans*, 100062.

Qi, H., Kim, M., Sandoval, C. L., Wang, Z., Schurgers, C., Lubarda, M. V., Baghdadchi, S., emily Gedney, X., Phan, A. M., & Delson, N. (2023). Board 400: The impact of Oral Exams on Engineering Students' Learning. 2023 ASEE Annual Conference & Exposition,

Reckinger, S. J., & Reckinger, S. M. (2022). A study of the effects of oral proficiency exams in introductory programming courses on underrepresented groups. Proceedings of the 53rd ACM Technical Symposium on Computer Science Education-Volume 1,

Ringle, C. M., Sarstedt, M., Mitchell, R., & Gudergan, S. P. (2020). Partial least squares structural equation modeling in HRM research. *The international journal of human resource management*, *31*(12), 1617-1643.

Roll, I., & Wylie, R. (2016). Evolution and revolution in artificial intelligence in education. *International journal of artificial intelligence in education*, *26*, 582-599.

Ryan, R. M., & Deci, E. L. (2000). Self-determination theory and the facilitation of intrinsic motivation, social development, and well-being. *American psychologist*, *55*(1), 68.

Sabin, M., Jin, K. H., & Smith, A. (2021). Oral exams in shift to remote learning. Proceedings of the 52nd ACM Technical Symposium on Computer Science Education,

Salehian Kia, F., Hatala, M., Baker, R. S., & Teasley, S. D. (2021). Measuring students' self-regulatory phases in LMS with behavior and real-time self report. LAK21: 11th international learning analytics and knowledge conference,

Touvron, H., Lavril, T., Izacard, G., Martinet, X., Lachaux, M.-A., Lacroix, T., Rozière, B., Goyal, N., Hambro, E., & Azhar, F. (2023). Llama: Open and efficient foundation language models. *arXiv preprint arXiv:2302.13971*.

Walter, Y. (2024). Embracing the future of Artificial Intelligence in the classroom: the relevance of AI literacy, prompt engineering, and critical thinking in modern education. *International Journal of Educational Technology in Higher Education*, *21*(1), 15.

Wang, B., Rau, P.-L. P., & Yuan, T. (2023). Measuring user competence in using artificial intelligence: validity and reliability of artificial intelligence literacy scale. *Behaviour & Information Technology*, *42*(9), 1324-1337.

Webb, N. L. (2002). Depth-of-knowledge levels for four content areas. *Language Arts*, *28*(March), 1-9.

Wu, R., & Yu, Z. (2024). Do AI chatbots improve students learning outcomes? Evidence from a meta-analysis. *British Journal of Educational Technology*, *55*(1), 10-33.

Xiao, J., Alibakhshi, G., Zamanpour, A., Zarei, M. A., Sherafat, S., & Behzadpoor, S.-F. (2024). How AI literacy affects students' educational attainment in online learning: testing a structural equation model in higher education context. *International Review of Research in Open and Distributed Learning*, *25*(3), 179-198.

Xu, Z. (2024). AI in education: Enhancing learning experiences and student outcomes. *Applied and Computational Engineering*, *51*(1), 104-111.

Yanai, I., & Lercher, M. J. (2024). It takes two to think. *Nature Biotechnology*, *42*(1), 18-19.

Yin, J., Goh, T.-T., Yang, B., & Xiaobin, Y. (2021). Conversation technology with micro-learning: The impact of chatbot-based learning on students' learning motivation and performance. *Journal of Educational Computing Research*, *59*(1), 154-177.

Zawacki-Richter, O., Marín, V. I., Bond, M., & Gouverneur, F. (2019). Systematic review of research on artificial intelligence applications in higher education–where are the educators? *International Journal of Educational Technology in Higher Education*, *16*(1), 1-27.

Zhang, Q., Nie, H., Fan, J., & Liu, H. (2025). Exploring the Dynamics of Artificial Intelligence Literacy on English as a Foreign Language Learners' Willingness to Communicate: The Critical Mediating Roles of Artificial Intelligence Learning Self-Efficacy and Classroom Anxiety. *Behavioral Sciences*, *15*(4), 523.

Zhang, W., Xiong, Y., Zhou, D., Liu, C., Gu, Y., & Yang, H. (2025). Balancing human and AI instruction: insights from secondary student satisfaction with AI-assisted learning. *Interactive Learning Environments*, 1-16.

# Supplementary Materials

*Table S1*

*Cronbach's Alpha and Composite Reliability*

| Construct | Cronbach's α | Composite Reliability |
|---|---|---|
| AI Ethics | 0.958 | 0.949 |
| AI Literacy | 0.957 | 0.964 |
| AI Self Competency | 0.893 | 0.902 |
| AI Self-Efficacy | 0.959 | 0.959 |
| Apply AI | 0.914 | 0.919 |
| Create AI | 0.982 | 0.982 |
| Detect AI | 0.920 | 0.927 |
| Effectiveness | 0.870 | 0.870 |
| Engagement | 0.878 | 0.879 |
| Satisfaction | 0.739 | 0.753 |
| Understand AI | 0.932 | 0.933 |
| Usability | 0.866 | 0.887 |

*Table S2*

*Average Variance Extracted (AVE)*

| Construct | AVE | t | p |
|---|---|---|---|
| AI Ethics | 0.905 | 74.259 | .000 |
| AI Literacy | 0.432 | 21.812 | .000 |
| AI Self Competency | 0.651 | 20.258 | .000 |
| AI Self-Efficacy | 0.829 | 49.952 | .000 |
| Apply AI | 0.701 | 21.871 | .000 |
| Create AI | 0.950 | 137.053 | .000 |
| Detect AI | 0.862 | 46.486 | .000 |
| Effectiveness | 0.658 | 17.074 | .000 |
| Engagement | 0.674 | 24.068 | .000 |
| Satisfaction | 0.792 | 28.608 | .000 |
| Understand AI | 0.748 | 36.377 | .000 |
| Usability | 0.659 | 15.323 | .000 |

*Table S3*

*Structural Path Coefficients*

| Path | β | t | p |
|---|---|---|---|
| AI Literacy → Effectiveness | 0.041 | 0.987 | .324 |
| AI Literacy → Engagement | 0.160 | 3.962 | .000 |
| AI Literacy → Satisfaction | 0.212 | 3.620 | .000 |
| AI Literacy → Usability | 0.301 | 5.093 | .000 |
| Apply AI → AI Literacy | 0.216 | 11.981 | .000 |
| Understand AI → AI Literacy | 0.264 | 23.441 | .000 |
| Create AI → AI Literacy | 0.145 | 8.935 | .000 |
| AI Ethics -> AI Literacy | 0.148 | 18.313 | .000 |
| AI Self-Efficacy → AI Literacy | 0.297 | 20.163 | .000 |
| Satisfaction → Effectiveness | 0.282 | 4.523 | .000 |
| Usability → Effectiveness | 0.608 | 8.200 | .000 |
| Usability → Engagement | 0.319 | 4.366 | .000 |
| Satisfaction → Engagement | 0.477 | 7.020 | .000 |

*Table S4*

***R-squared Values for Endogenous Constructs***

| Construct | R² | t | p |
| --- | --- | --- | --- |
| AI Literacy | 1.000 | 27380.01 | .000 |
| Effectiveness | 0.616 | 10.76 | .000 |
| Engagement | 0.547 | 10.85 | .000 |
| Satisfaction | 0.045 | 1.755 | .079 |
| Usability | 0.091 | 2.510 | .012 |

*Table S5*

***Effect Sizes ($f^2$)***

| Path | $f^2$ | t | p |
| --- | --- | --- | --- |
| AI Self-Efficacy → AI Literacy | 1205.19 | 2.671 | .008 |
| Apply AI → AI Literacy | 1099.70 | 2.616 | .009 |
| Understand AI → AI Literacy | 748.56 | 2.600 | .009 |
| Usability → Effectiveness | 0.743 | 2.548 | .011 |
| Satisfaction → Effectiveness | 0.168 | 2.311 | .021 |
| Satisfaction → Engagement | 0.408 | 2.917 | .004 |
| AI Literacy → Usability | 0.100 | 2.193 | .028 |

*Table S6*
*Group Differences by Gender*

| Measure | Group | M | SD | t(df) | p |
|---|---|---|---|---|---|
| Average Usability | Male | 4.65 | 0.58 | -0.54(301) | .590 |
|  | Female | 4.69 | 0.45 |  |  |
| Average Effectiveness | Male | 4.59 | 0.58 | -1.19(301) | .235 |
|  | Female | 4.66 | 0.47 |  |  |
| Average Engagement | Male | 4.31 | 0.73 | -1.28(301) | .203 |
|  | Female | 4.41 | 0.67 |  |  |
| AI Literacy | Male | 8.66 | 1.41 | 3.21(301) | <.001 |
|  | Female | 8.13 | 1.46 |  |  |
| AI Exposure | Male | 2.59 | 1.05 | 1.45(293) | .149 |
|  | Female | 2.42 | 0.97 |  |  |
| Satisfaction | Male | 4.30 | 0.70 | -1.28(301) | .200 |
|  | Female | 4.41 | 0.71 |  |  |
| Apply AI | Male | 9.51 | 1.48 | 0.09(301) | .930 |
|  | Female | 9.50 | 1.59 |  |  |
| Understand AI | Male | 8.92 | 1.64 | 3.22(297) | <.001 |
|  | Female | 8.26 | 1.89 |  |  |
| Detect AI | Male | 8.94 | 1.66 | 1.75(300) | .082 |
|  | Female | 8.59 | 1.80 |  |  |
| Create AI | Male | 6.50 | 3.00 | 3.42(300) | <.001 |
|  | Female | 5.33 | 2.90 |  |  |
| AI Self-Efficacy | Male | 8.63 | 2.07 | 3.03(300) | .003 |
|  | Female | 7.90 | 2.09 |  |  |
| AI Self-Competency | Male | 9.45 | 1.37 | 1.67(301) | .095 |
|  | Female | 9.15 | 1.66 |  |  |